\documentclass[reprint]{aastex62}

\received{12 April 2019}
\accepted{23 December 2019 by {\bf The Astrophysical Journal}}

\shorttitle{Transmission spectra of hot-Jupiters}
\shortauthors{Sengupta, Chakrabarty and Tinetti}

\begin{document}

\title{OPTICAL TRANSMISSION SPECTRA OF HOT-JUPITERS: EFFECTS OF SCATTERING }

\correspondingauthor{Sujan Sengupta}
\email{sujan@iiap.res.in}

\author{Sujan Sengupta}
\affil{Indian Institute of Astrophysics, 
Koramangala 2nd Block, Sarjapura Road, 
Bangalore 560034, India}

\author{Aritra Chakrabarty}
\affil{Indian Institute of Astrophysics, Koramangala 2nd Block,
Sarjapura Road, Bangalore 560034, India}
\affil{University of Calcutta, Salt Lake City, JD-2 
Kolkata 750098, India}

\author{Giovanna Tinetti}
\affiliation{University College London, 
Dept. of Physics and Astronomy, 
Gower Street, London, UK WC1E6BT}

\begin{abstract}
 We present new grids of transmission spectra for hot-Jupiters by solving
the multiple scattering radiative transfer equations with non-zero scattering
albedo instead of using the Beer-Bouguer-Lambert law for the change in the
transmitted stellar intensity. The diffused reflection and transmission due
to scattering  increases the transmitted stellar flux resulting into a decrease
in the transmission depth. Thus we demonstrate that scattering plays a double role in
determining the optical transmission spectra -- increasing the total optical
depth of the medium  and  adding the diffused radiation due to scattering to the
transmitted stellar radiation. The resulting effects
yield into an increase in the transmitted flux and hence reduction  
in the transmission depth. For a cloudless planetary atmosphere, Rayleigh scattering
albedo alters the transmission depth up to about 0.6 micron but the change in the
transmission depth due to forward scattering by cloud or haze is significant throughout the
optical and near-infrared regions. However, at wavelength longer than about 
1.2 $\mu$m, the scattering albedo becomes negligible
and hence the transmission spectra match with that calculated without solving
the radiative transfer equations.  We compare our model spectra with existing
theoretical models and find significant difference at wavelength shorter than
one micron. We also compare our models with observational data for a few hot-Jupiters
which may help constructing better retrieval models in future.

\end{abstract}


\section{Introduction} \label{sec:intro}

While planetary transit photometry provides important physical properties
of exoplanets, it cannot explore the planetary atmosphere. As pointed out
for the first time by \cite{seager00}, it is the transmission spectroscopic
method that can probe the physical and chemical properties of the atmosphere of
exoplanets having near edge-on orientation. 

  During the transit epoch of an exoplanet across its parent star, a part of
the starlight passes through the planetary atmosphere. The interaction of this
star-light with the atmospheric material through absorption and scattering
is imprinted on top of the stellar spectra. A correct interpretation of this
transmission spectra needs a comparison with a consistent theoretical model 
that incorporates all the physical and chemical processes in the planetary
atmosphere.         

   Theoretical models for transmission spectra of stars with transiting
exoplanets having a wide range of equilibrium temperature and surface gravity
have already been presented by several groups, e.g. \citep{brown01, tinetti07, 
madhu09, burrows10, fortney10, griffith14, waldmann15,
kempton17, barstow17, heng18, goyal18,goyal19}. Useful review and overview 
on modeling exoplanetary atmosphere can be found in \cite{fortney18,burrows14,
tinetti13}. 

The models by  \cite{fortney10} and \cite{kempton17} are based on 
thermo-chemical equilibrium scheme, use elemental abundances from 
\cite{lodders03} and atomic-molecular line-list primarily from HITRAN
\citep{gordon17}. On the other hand, model by \cite{goyal19,goyal18} 
uses elemental abundances from \cite{asplund09} and the line-list 
from Exomol \citep{tennyson16}. Nevertheless all the three models
agrees well at low temperature.     

  In all these three models and the other models mentioned above, only
absorption of starlight passing through the planetary atmosphere is 
incorporated and thus the reduced intensity $I$ due to the interaction of
atoms and molecules in the atmosphere is calculated by using the
Beer-Bouguer-Lambert law $I=I_0e^{-\tau}$ where $I_0$ is the incident stellar
intensity and $\tau$ is the line-of-sight optical depth of the medium that
imprint the signature of the planetary atmosphere. In these models, although
opacity due to scattering is added up to the opacity due to true absorption,
angular distribution of the transmitting photon due to scattering is not 
incorporated. Since scattering co-efficient and hence single scattering albedo
at longer wavelengths, e.g., in infrared is extremely small or zero, this
approximation is valid at wavelengths
beyond the optical region. But it overestimates the transmission 
depth at shorter wavelength and hence does not provide correct results for 
optical region where scattering albedo is comparable to 1 and the diffused
transmission and reflection due to scattering plays important role in 
determining the radiation  field. A correct treatment is thus to solve the 
multi-scattering radiative transfer equations for the diffused reflection 
and transmission as demonstrated by \cite{stam2012} who presented 
three dimensional Monte Carlo simulation for Titan's atmosphere at wavelengths ranging
between 2.0 and 2.8 micron and reported significant underestimation in the 
calculation of the transmission flux if forward scattering by haze and gas is
neglected in the retrieval models. 

  In this paper we present the transmission depth as the solution of the
detail multiple-scattering radiative transfer equations for the atmosphere of
exoplanets with a wide range of equilibrium  temperature and surface gravity.

Today a few tens of gaseous exoplanet atmospheres have been probed in the
optical and  near-IR through transit observations with the Wide Field Camera 3 
on board Hubble Space Telescope \citep{tsiaras18}. 
For a sub-sample of those, also optical spectra
using the Space Telescope Imaging Spectrograph are available \citep{sing16}.
This survey was complemented by photometric transit observations at two more
longer wavelengths 3.6 $\mu m$ and 4.5 $\mu m$ by using the Spitzer Space
Telescope Infrared Array Camera. Although one needs to be careful in combining
data from multiple instruments \citep{yip19}, these observational data provide an excellent
opportunity to understand the scope and limitations of various theoretical
models. 

We compare our model spectra with the existing theoretical models and with
the observed HST and Spitzer data. In the next section we provide the
formalisms  for calculating the transmission depth. In Section~3 we discuss
the model absorption and scattering opacity adopted in our present models.
Section~4 outlines the numerical method for solving the multiple scattering
radiative transfer equations. The non-isothermal temperature-pressure profiles
for non-Grey planetary atmosphere used in our models are described in 
Section~5. In Section~6 we present a simple haze model that is incorporated
in order to include additional absorption and scattering opacities. The
results are discussed in Section~7 followed by specific conclusion in the
last section.
     
\section{The Transmission Depth} \label{sec: transdpt}

  The transmission spectra of exoplanets are expressed in term of the
wavelength dependent transmission depth which is given by \cite{kempton17}

\begin{eqnarray}
D_{\lambda}=1-\frac{F_{in}}{F_{out}},
\end{eqnarray}\label{tdepth1}
where $F_{out}=F_{\star}$ is the out-of-transit stellar flux. The in-transit
stellar flux $F_{in}$ which is the flux of the host star that transmits
through the planetary atmosphere is given by
\begin{eqnarray}
F_{in}=\left(1-\frac{R_{PA}^2}{R_{\star}^2}\right)F_{\star}+F_P,
\end{eqnarray}\label{tdepth2}
where $R_{PA}$ is the combined base radius $R_{P}$ of the planet and 
its atmosphere, $R_{\star}$ is the radius of the host star and $F_P$ is the
additional stellar flux that passes through the planetary atmosphere and
suffers absorption and scattering.
Clearly, the first term  in the right hand side of the above
expression represents the stellar radiation during the transit of the planet 
and its atmosphere and the second term
represents the additional stellar radiation filtered through the planetary atmosphere.
The base radius $R_P$ is the planetary radius at which the planet
becomes opaque at all wavelength. For a rocky planet, $R_P$ is the distance
between the center to the planetary surface. But for gaseous planets, $R_P$
is the height of the region bellow which no radiation can transmit from.      

From  the above equations, the transmission depth can be written in a simple
form 
\begin{eqnarray}
D_{\lambda}=\frac{R_{PA}^2}{R_{\star}^2}-\frac{F_P}{F_{\star}}.
\end{eqnarray}

Sometimes the transmission spectra is expressed in terms of the wavelength
dependent planet-to-star radius ratio which is the square root of 
$D_{\lambda}$.

 The stellar radiation $F_P$ that filters through the planetary atmosphere 
is calculated from the incident stellar intensity.
If the calculations of transmission spectra assumes only absorption of starlight passing
through the planetary atmosphere, Beer-Bouguer-Lambert law  can be used which is given by

\begin{eqnarray}
I(\lambda)=I_0(\lambda)e^{-\tau_{\lambda}/\mu_0},
\end{eqnarray}\label{bbl}
where $I_0$ is the intensity of the incident stellar radiation, $I$ is the
stellar intensity filtered through the planetary atmosphere, $\tau$ is the
optical depth along the ray path and $\mu_0$ is the cosine of the angle 
between the direction of the incident starlight and the normal to the planetary
surface. Due to the edge on orientation, $\mu_0=1$ is adopted in the present
investigation such that the starlight during planetary transit always incident 
along the normal to the planetary atmosphere.

Although in many previous models, opacity due to scattering $\sigma$ is 
added to the true absorption $\kappa$, scattering into and out of the ray is
not explicitly considered before. This assumption is reasonable
for calculating the transmission spectra at longer wavelength, e.g., in the
infra-red region where the scattering albedo is negligible. 
But it grossly overestimate the transmission depth and hence
does not provide correct results for optical region where scattering albedo
$\omega$ which is the ratio of the scattering co-efficient to the extinction
coefficient is non-zero and plays important role in determining the radiation
field. It's worth mentioning that $\omega$ depends on the wavelength as well
as the atmospheric depth. A true treatment is thus to solve the multi-scattering radiative 
transfer equations for diffused reflection and transmission which for a
 plane-parallel geometry is given by \cite{chandra60}     

\begin{eqnarray}
\mu\frac{dI(\tau,\mu,\lambda)}{d\tau}=I(\tau,\mu,\lambda)-
\frac{\omega}{2}\int_{-1}^1{p(\mu,\mu')I(\tau,\mu',\lambda)\text{d}\mu'}
-\frac{\omega}{4}F e^{-\tau/\mu_0}p(\mu,\mu_0),
\end{eqnarray}
where $I(\tau,\mu,\lambda)$ is the specific intensity of the diffused radiation
field  along the direction $\mu=\cos\theta$, $\theta$ being the angle between 
the axis of symmetry and the
ray path, $F$ is the incident stellar flux in the direction $-\mu_0$,
 $\omega$ is the 
albedo for single scattering, $p(\mu,\mu')$ is the 
scattering phase function that describes the angular distribution of the
photon before and after scattering and $\tau$ is the optical depth along
the line of sight given by \cite{tinetti13} 
\begin{eqnarray}
\tau(\lambda,z)=2\int_0^{l(z)}{\chi(\lambda,z)\rho(z) \text{d}l}.
\end{eqnarray}
In the above equation $\chi$ is the extinction co-efficient which is 
the sum of the absorption coefficient $\kappa$ and scattering co-efficient $\sigma$,
$\rho(z)$ is the atmospheric density, $z$ is the atmospheric height along 
the axis of symmetry of the planet and $l$ is the path traveled by the
stellar photon and can be written as \citep{tinetti13}   
\begin{eqnarray}
l(z)=\int{\text{d}l}=\sqrt{(R_P+z_{max})^2-(R_P+z)^2},
\end{eqnarray}
$z_{max}$ is the atmospheric height above which the stellar photon does not
suffer any scattering or absorption.

The scattering phase function depends on the nature of scatterers. For 
scattering by non-relativistic electrons (Thomson scattering) and by atoms
and molecules, the angular distribution is described by Rayleigh scattering
phase function and is given by \citep{chandra60}
\begin{eqnarray}
p(\mu,mu')=\frac{3}{4}[1+\mu^2\mu'^2+\frac{1}{2}(1-\mu^2)(1-\mu'^2)],
\end{eqnarray}   
where $\mu$ and $\mu'$ are the cosine of the angle before and after scattering
with respect to the normal. 

A beam of radiation traversing in a medium  gets weakened by its interaction 
with matter by an amount $\text{d}I_\nu=-k_\nu\rho I_\nu \text{d}s$=$-I_\nu \text{d}\tau$ where 
$\rho$ is the density of the medium and $\kappa_\nu$ is the mass absorption 
co-efficient. Integration of this expression yields into the 
Beer-Bouguer-Lambert law.
  As pointed out by \cite{chandra60}, while passing through a medium,
this reduction in intensity suffered by a beam  of radiation is not necessarily
lost to the radiation field. 
A fraction of the energy lost from an incident beam  would reappear in other
directions due to scattering and the remaining part would have been truly
absorbed in the sense that it may get transformed into other form of energy
or of radiation of different frequencies. For a scattering atmosphere, the 
scattered radiation from all other directions contribute to the 
emission co-efficients into the beam of the direction considered.  

In a scattering medium, the radiation field has two components: the reflected
and the transmitted intensities which suffer one or  more scattering processes
and the directly transmitted flux $\pi Fe^{-\tau/\mu_0}$ in the direction 
$-\mu_0$. So, the reflected and the transmitted intensities that is 
incorporated through the second term in the right hand side of 
Equation~5, does not include the directly transmitted flux which is
described by the third term. In other words, the reduced incident
radiation $\pi F e^{-\tau/\mu_0}$ which penetrates to the atmospheric level
$\tau$ without suffering any scattering is different than
the diffuse radiation field $I(\tau,\mu)$ which has arisen because of one or
more scattering processes.  Therefore, in the absence of scattering, 
i.e., when $\omega=0$, the emergent intensity obtained by integrating 
Equation~5 reduces to that given by Beer-Bouguer-Lambert law.  
As a consequence, in the infra-red wavelength region where the scattering albedo is negligibly 
small or zero, use of Beer-Bouguer-Lambert law $I=I_0e^{-\tau}$ in calculating
the transmission depth is appropriate.

Solution of the above radiative transfer equation provides the intensity along
the direction $\mu$ of the stellar radiation that passes through the planetary
atmosphere.  The reduced stellar flux $F_P$ that emerges out of the planetary 
atmosphere is obtained by integrating the intensity
in each beam of radiation, over the solid angle subtended by the atmosphere.  

\section{The absorption and scattering opacity} \label{sec: opacity}

The main aim of the present work is to calculate the transmission spectra
appropriate in the optical wavelength region. We do not intend to investigate
the chemistry under different conditions of the atmosphere. Therefore 
we present models with a fixed
metallicity - solar metallicity and solar system abundances for the atoms and
molecules in the planetary atmosphere. We calculate the gas absorption and 
scattering co-efficients by using the software package "Exo-Transmit" 
\citep{kempton17} available in the public domain\footnote{
${\rm https://github.com/elizakempton/Exo_Transmit}$}. The molecular opacities
are adopted from the well-known and well-used data-base of 
\cite{freedman08,freedman14}. In Exo-Transmit software package, the equation
of states (EOS) of various species that provides the abundances for major
atmospheric constituents as a function of temperature and pressure are 
calculated based on the solar system abundances of \cite{lodders03}. The 
abundances of all the species are in chemical equilibrium and the EOS for all
atomic and molecular species are computed for a temperature range of 100-3000K
and for a pressure range of $10^{-9}-1000$ bars. Opacities for 28 molecular
species as well as Na and K are tabulated in a fixed $T-P$ grid for
wavelengths ranging from 0.3 $\mu {\rm m}$ to 30 $\mu {\rm m}$ at a fixed
spectral resolution of 1000. The line
list used to generate the molecular opacity is tabulated in \cite{lupu14}.
The collision-induced opacities weighted by the product of the abundances of
the pair of molecules along with the Rayleigh scattering opacity is added to
the sum of the individual opacities of all the molecular and atomic opacities
weighted by their abundances for each temperature-pressure-wavelength points.    

\begin{figure}[t]
\includegraphics[angle=0,scale=0.9]{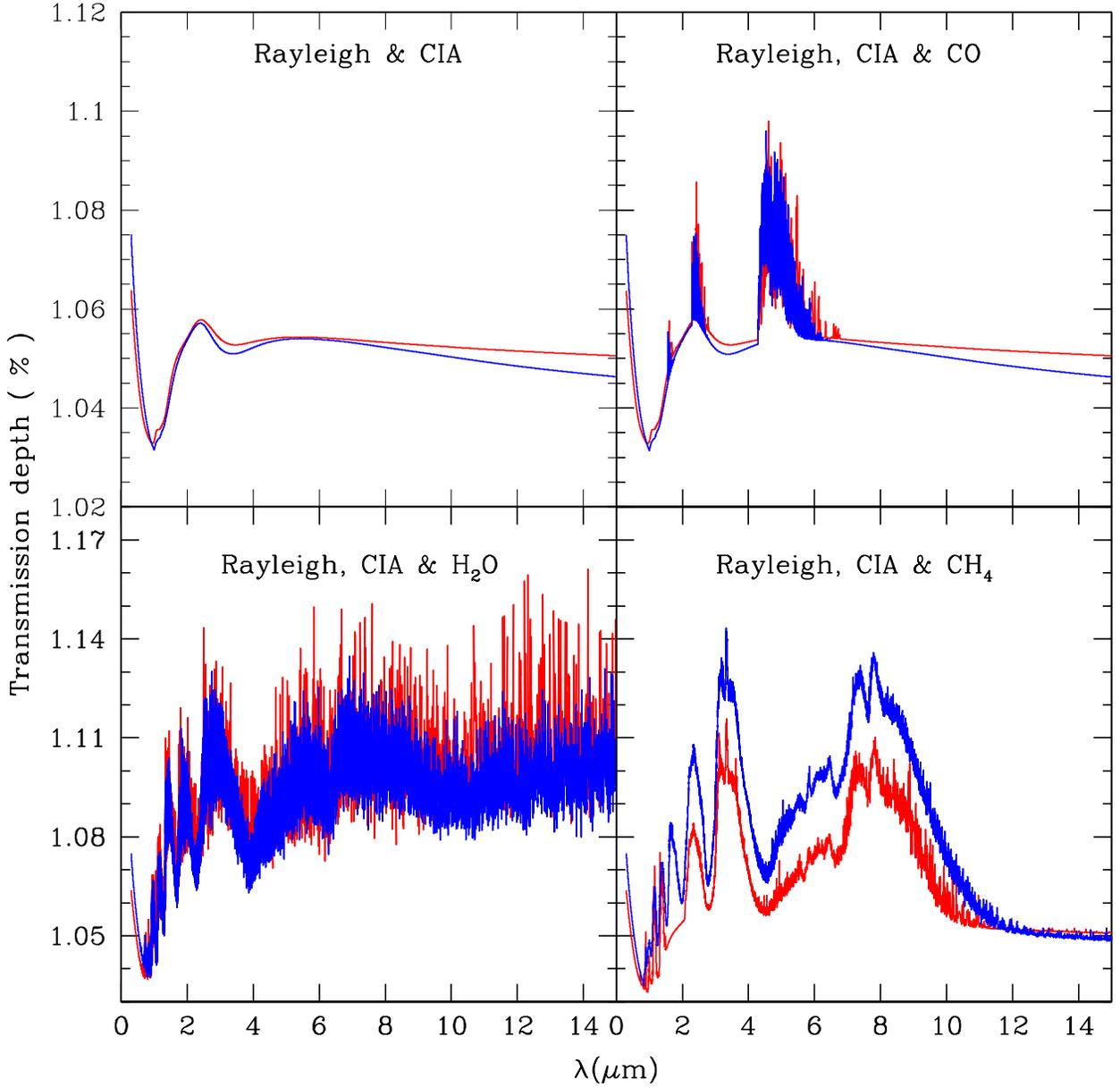}
\caption{Comparison of transmission depth calculated by using two  different
models : (1) Exo-Transmit (red) and (2) Tau-REx (blue). A  Jupiter-size planet
with $T_{eq}=2700K$ and ${\rm g=30 ms^{-2}}$ transiting a star with solar
radius is considered in calculating the transmission depth.}    
\end{figure}

In order to investigate the consistency and correctness in the chemistry
involved, we derived the opacity due to a few individual molecules along with 
the opacity due to Rayleigh scattering and Collisional Induced Absorption (CIA)
by using another  model, Tau-REx \citep{waldmann15}. Tau-REx is an open source model which
adopts line-lists from ExoMol \citep{tennyson12, tennyson16}.
By using the optical depth derived from the two different models, 
we calculate and compare the transmission depth for a jupiter 
size exoplanet with $T_{eq}= 2700K$ and surface gravity ${\rm g=30 ms^{-2}}$
transiting a solar type star. We set the abundances of the individual
molecules CO, ${\rm H_2O}$ and ${\rm CH_4}$ at $10^{-4}$ and used the 
Beer-Bouguer-Lambert law and not the radiative transfer equations for
calculating the transmission depth. Figure~1
shows that except for ${\rm CH_4}$, the results obtained by using the opacity
derived from  the two models are in good agreement. We will address this
difference in a future paper as we limit the scope of the present work
to the effect of scattering in the optical.     

Using Exo-Transmit package we calculated the total extinction co-efficients
(true absorption plus scattering) as well as the scattering 
co-efficients for a given $T-P$ profile and surface gravity. The albedo
for single scattering at each wavelength and each pressure point is
calculated  by taking the ratio of the scattering co-efficient and the 
extinction co-efficient. We have incorporated all the species provided
in the package and their EOS for solar metalicity without any change.
The EOS for Rain-out condensation are adopted in all the calculations. 

   Finally, we have not included cloud opacity or additional scattering
sources in our use of Exo-Transmit package. We have incorporated haze
in our radiative transfer code and we discuss the cloud model in 
section~\ref{sec: cloud}. The Exo-Transmit software package is used only to
calculate the atomic and molecular absorption and scattering coefficients.

\section{Numerical method to solve the radiative transfer equations}

  We use the absorption and scattering co-efficients at different pressure level
in the planetary atmosphere and calculate the line of sight optical depth as 
given in Equation~6. The wavelength dependent albedo for single
scattering $\omega$ at different pressure levels is the ratio between the 
scattering co-efficients $\sigma(\lambda)$ and the extinction co-efficient
$\chi(\lambda)$. We solve the multiple scattering radiative transfer equation
as given in Equation~5 by using discrete space theory developed by
\cite{peraiah73}. The numerical code is extensively used to solve the vector 
radiative transfer equations in order to calculate polarized spectra of cloudy 
brown dwarfs and self-luminous exoplanets \citep{sengupta09,sengupta10,
marley11,sengupta16,sengupta16a,sengupta18}. For the present work we use
the scalar version  of the same numerical code.

In this method we adopt the following steps :

\begin{enumerate}
\item  The medium is divided into a number of ``cells'' whose thickness is defined
by $\tau$. The thickness of each cell is less than a critical optical thickness
$\tau_c$ which is determined on the basis of the physical characteristics of
the medium .
\item  The integration of the radiative transfer equation is performed on the cell
which is bounded by a two dimensional grids $[\tau_n,\tau_{n+1}]\times
[\mu_{j-1/2},\mu_{j+1/2}]$.
\item  These discrete equations are compared with the canonical equations of the
interaction principle and the transmission and reflection operators of cells
are obtained.
\item  Lastly, all the cells are combined by ``star'' algorithm and the radiation
field is obtained.
\end{enumerate}
A detail description of the numerical method can be found in \cite{peraiah73,
sengupta09}.   

 Using 2.5 GHz Intel core i5 processor with 8 GB RAM, it takes typically 10-12 
minutes for one complete run of the FORTRAN version of the code that calculates the
transmission spectra for wavelength ranging from 0.3-30 $\mu{\rm m }$ with a total number of 
4616 wavelength points. We have also developed python version of the code which provides the
same results in shorter time.    

In order to validate the numerical method as well as the molecular and atomic
opacity used in the present work, we present in Figure~2 a comparison of our
model spectrum with a model by \cite{stephens09} for a cloud-free methane-dwarf
(T8) and with the observed Spex prism  spectrum  \citep{burg04} of the T-dwarf
2MASSI J0415-0935. We also compare our model spectrum with the model presented 
by \cite{fortney08} for a self-luminous directly imaged Jupiter-type exoplanets
with $T_{eff}=600$K and surface gravity $g=30 {\rm ms^{-2}}$. The comparison
is presented in Figure~3. The model spectra and the 
temperature-pressure profiles for both the cases have kindly been provided
by M. Marley (private communication).   

\begin{figure}[]
\includegraphics[angle=-90,scale=0.7]{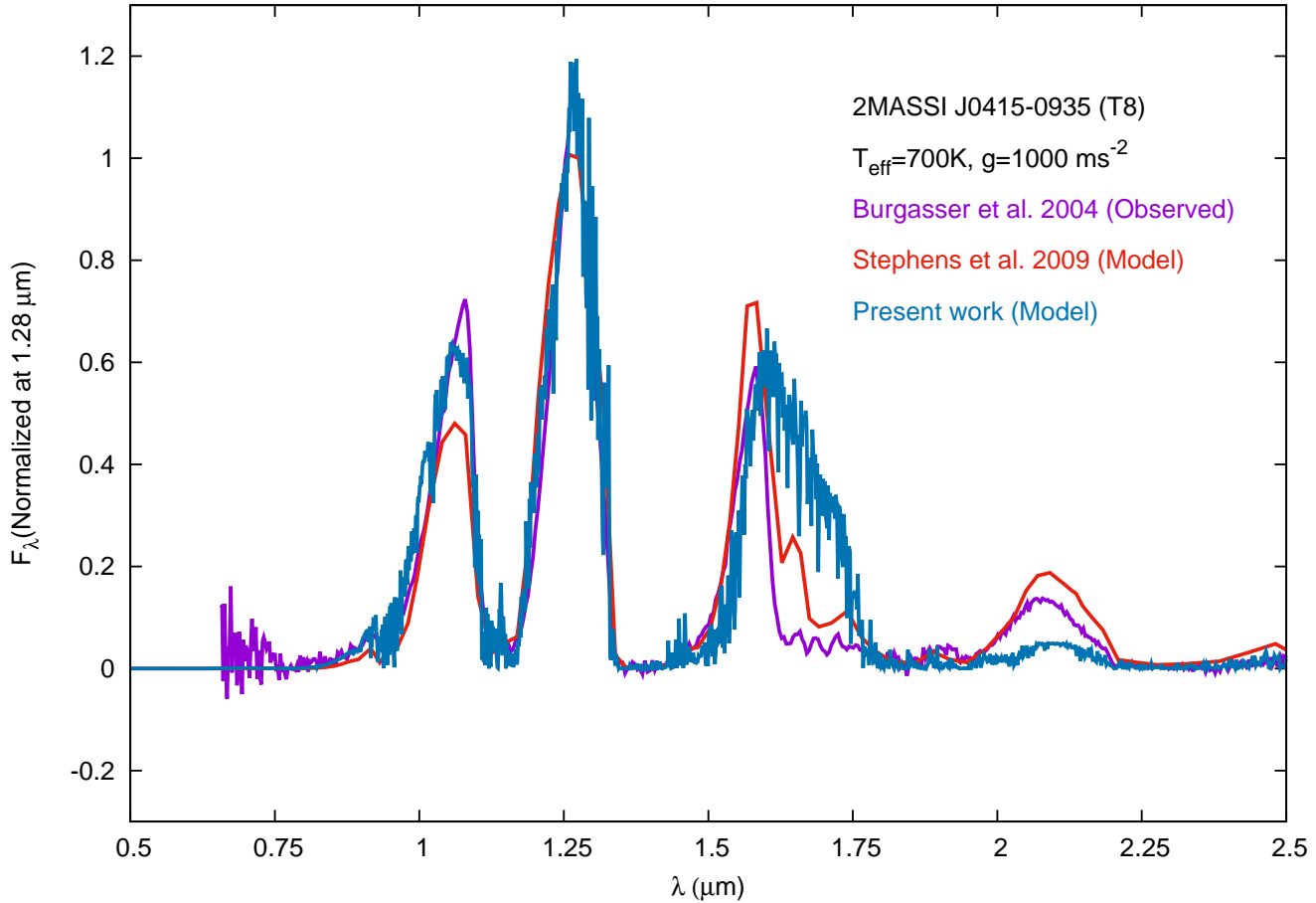}
\caption{Comparison of model spectra with the observed Spex prism  spectrum
of a cloud-free brown dwarf (T8) 2MASSI J0415-0935.\label{fig:fig2}}
\end{figure}

\begin{figure}[]
\includegraphics[angle=-90,scale=0.7]{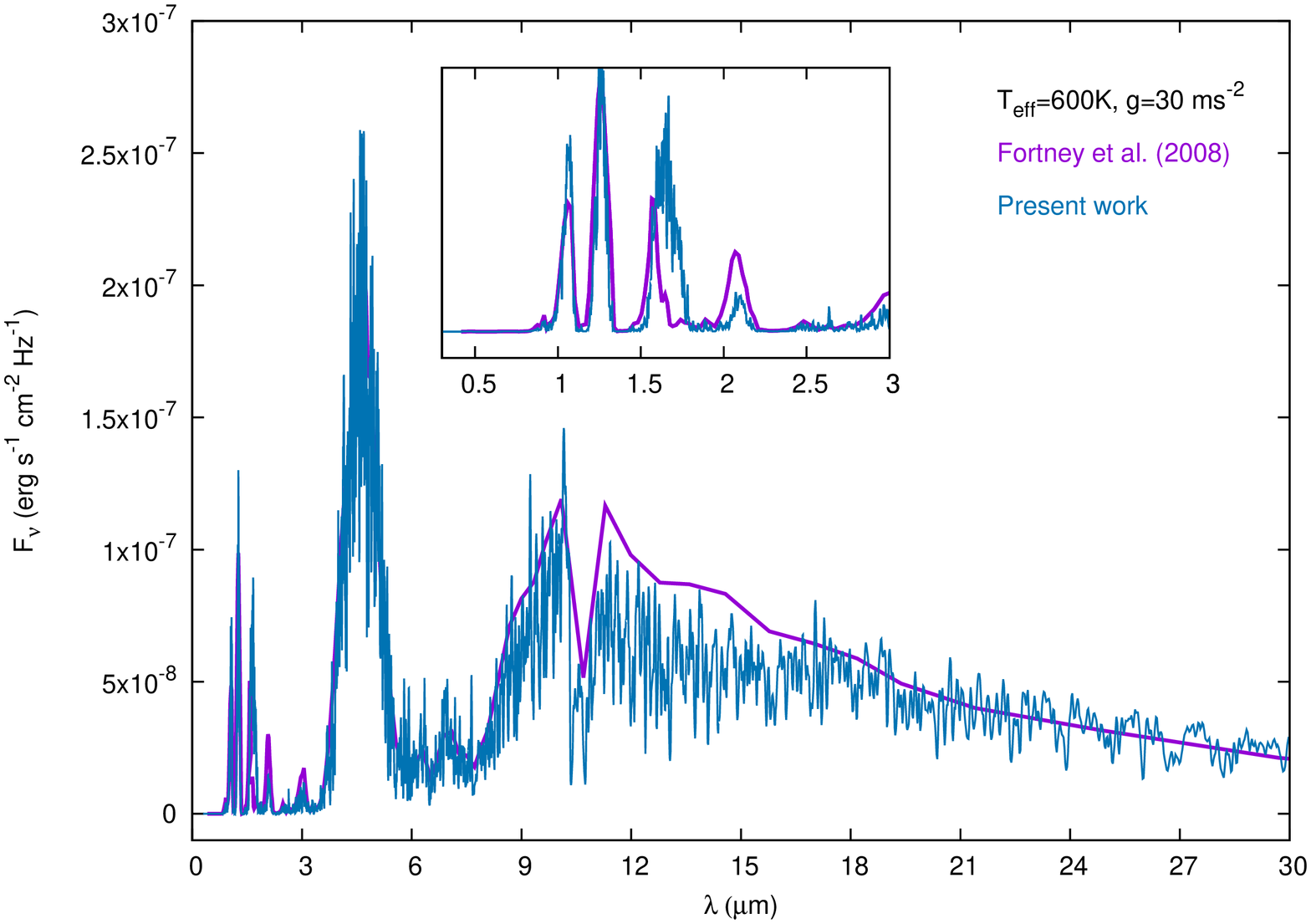}
\caption{Comparison of model spectra for a self-luminous giant exoplanet}
\end{figure}

 The slight miss-match of our synthetic spectrum with that of \cite{stephens09} at the
infra-red region of the T-dwarf, as presented in Figure~2, is due to the disagreement in the
opacity of methane as detected while comparing the model transmission depth derived by 
using Exo-Transmit and Tau-REx. The difference may also be attributed to a different
elemental abundaces adopted. It is worth mentioning here that for the case of self-luminous
exoplanet, the model spectrum of \cite{fortney08} incorporates condensate cloud in the visible
atmosphere while we have considered a cloud-free atmosphere.         

\section{The temperature-pressure profiles for irradiated exoplanets} \label{sec:tpprofile}

The atmospheric temperature structure is an important input in the calculation
of the transmission spectra. The self-consistent way to obtain the
temperature-pressure ($T-P$) profile is to solve the radiative equilibrium
equations simultaneously with the radiative transfer equations and hydrostatic
equilibrium equations. The presence of molecules makes it more difficult
to estimate the temperature structure as chemical equilibrium equations too
need to be solved self-consistently. Further, for strongly irradiated
exoplanets, the internal temperature is negligible compared to the temperature
due to irradiation and the incident stellar flux at the top-most layer 
determines the atmospheric temperature structure as it interacts with the 
medium through absorption and scattering. Therefore, the atmospheric 
temperature at different depth is determined by the optical depth of the
medium.. At the same time, the optical depth is governed by the temperature
structure making it an involved and complicated numerical procedure. For 
stars, brown dwarfs and self-luminous exoplanets with weak or negligible 
irradiation, analytical formula for the $T-P$ profile in Grey or ``slightly''
non-Grey atmosphere was derived by \cite{chandra60}. Analytical formalisms
of temperature structure for non-Grey strongly irradiated planets are 
presented by \cite{hansen08,guillot10,parmentier14,parmentier15}. In order
to model the transmission spectra of close-in exoplanets, isothermal 
$T-P$ profiles with $T(P)=T_{eq}$ were adopted by \cite{sing16,kempton17,goyal19}.
The radially inwards incident radiation usually penetrates quite deep, about 10-100bars 
pressure level. However, in the transit geometry considered for calculating
the transmission spectrum, the atmosphere below approximately 1 bar pressure level is opaque 
because of the large path length that the radiation traverses. 
Therefore, a very small part of the overall atmosphere is probed in the transmission 
spectrum. So, isothermal approximation although not completely accurate, 
especially for hotter planets where temperature inversion due to the presence of
TiO and VO becomes dominant, does not make much difference in the results for 
the comparatively cooler planets \citep{goyal18} with current observations.  

In the present work, we have used the FORTRAN implementation of the analytical
model for the $T-P$ profiles of non-Grey irradiated planets presented by \cite
{parmentier14, parmentier15}. This code available in public domain\footnote{$
{\rm http://cdsarc.u-strasbg.fr/viz-bin/qcat?J/A+A/574/A35}$} uses the
functional form  for Rosseland opacity provided by \cite{valencia13} which is
based on the Rosseland opacities of \cite{freedman08}. The analytical model
takes into account the opacities both in the optical and in the infra-red 
region. The analytical models are compared with the state-of-the-art 
numerical models and the different coefficients in the analytical models 
are calibrated for a wide range of surface gravity and equilibrium temperature.

\begin{figure}[]
\includegraphics[angle=-90,scale=0.7]{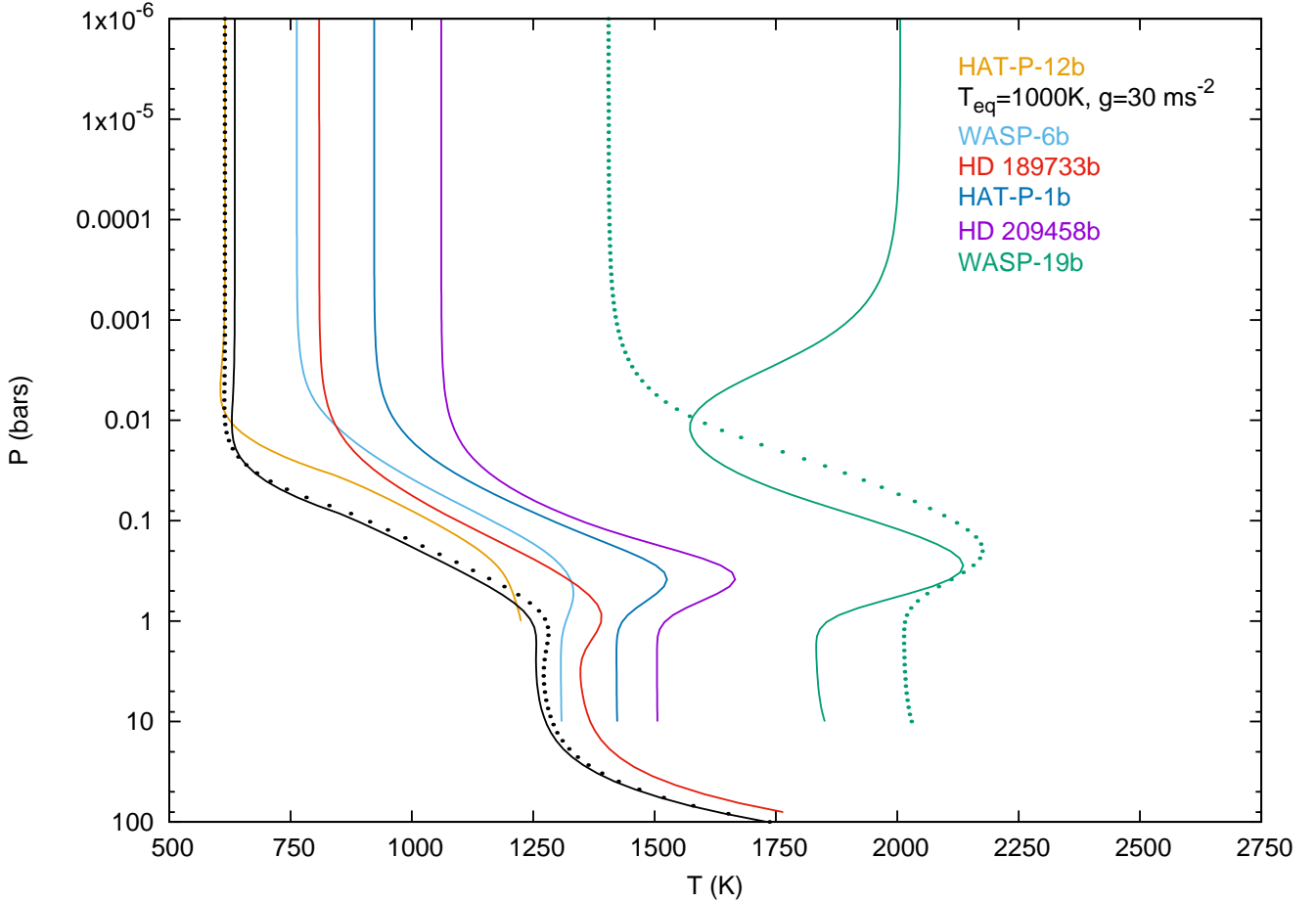}
\caption{ Temperature-Pressure ($T-P$) profiles of hot-Jupiters derived from
the analytical formalisms presented by \cite{parmentier15}. Dotted lines
represent the corresponding $T-P$ profiles without the presence of TiO and VO
in the atmosphere. $T_{eq}$ and $g$ used for each planet are listed in
Table~1.\label{fig:fig4}}
\end{figure}

In Figure~4 we present the $T-P$ profiles derived by using the above mentioned 
computer code for a number of exoplanets with a wide range of surface gravity
$g$ and equilibrium temperature $T_{eq}$. The values of $g$ and $T_{eq}$ for 
for various exoplanets are given in Table~1. We assumed that  
$T_{eq}$ for each exoplanets is the equilibrium temperature for zero albedo and
the thermal profile is planet-averaged (see \cite{parmentier14}). Solar flux is
assumed in the calculations. We have not
considered a convective zone at the bottom of the models as such a zone should
be situated much below the pressure level corresponding to $R_P$. 

\begin{table}
\begin{center}
\caption{Best fit parameters for the model transmission spectra in the infrared region
of six exoplanets}
\begin{tabular}{ccccccc}
\tableline\tableline
Name & $T_{eq}$ (K)  & g (${\rm ms^{-2}}$)  &
${\rm R_P (R_J)}$  & ${\rm  R_* (R_\odot)}$ &  $n_0$ & $d_0$ ($\mu$m )   \\
\tableline
WASP-19b & 2050 & 14.2 & 1.34 & 1.01 & 0.0 & 0.0  \\ 
HD 209458 & 1448 & 9.4 & 1.38 & 1.2 & $5\times10^4$  & 0.4 \\
HAT-P-1b & 1320 & 7.5 & 1.33 & 1.195 & $3\times10^4$ & 0.4 \\
HD 189733 & 1200 & 21.4 & 1.19 & 0.8 & $2\times10^5$ & 0.4 \\
HAT-P-12b & 960 & 5.6 & 0.9 & 0.71 &  $1\times10^6$ & 0.2 \\
WASP-6b & 1150 & 8.7 & 1.18 & 0.87 & $2\times10^5$ & 0.4 \\
\tableline
\end{tabular}
\end{center}
\end{table}

  We have included the effect of TiO and VO on the $T-P$ profile. As shown in
Figure~4, the effect is not significant for planets with $T_{eq}\le1000$K.
However, as $T_{eq}$ increases, the atmospheric temperature increases
significantly in the upper atmosphere due to the presence of TiO and VO 
in the atmosphere. For a planet as hot as WASP-19b ($T_{eq}$=2050K), 
the presence of TiO and VO introduces temperature inversion which disappears
in the absence of TiO and VO. It's worth mentioning here that in the absence 
of internal energy of the planet, the $T-P$ profile is used only to calculate
the absorption and scattering co-efficients. The radiation field is determined
by the incident stellar flux at the uppermost boundary.

\section{Additional absorption and scattering due to atmospheric cloud/haze}
\label{sec: cloud}

Condensation cloud may play important role in shaping the transmission spectra
of hot-Jupiters \citep{fortney10,sing16}. Under appropriate combination 
of temperature and surface gravity and based on chemical equilibrium process,
cloud or haze may form in the visible region of the
planetary atmosphere \citep{sudarsky03}. It is well-known that the
optical spectra of hotter L-brown dwarfs are shaped by the presence of 
condensation cloud \citep{cushing08} while in comparatively cooler T-brown 
dwarfs, cloud gets rain down below the visible atmosphere. Detail models 
of cloud and haze under chemical equilibrium
in exo-planetary atmospheres have been presented by \cite{ackerman01,cooper03,
burrows08}.

In the present work we have considered a simple model for thin haze in the
uppermost atmosphere following the approach of \cite{griffith98,saumon00}.
In this model the dust absorption and scattering cross-sections as well as the
scattering phase function are calculated
with the Mie theory of scattering \citep{bohren83}. The cloud is confined 
within a thin region of the atmosphere bound by a base and deck. The vertical
density distribution of the cloud particle is given by
\begin{eqnarray}
n(P)=n_0\frac{P}{P_0},
\end{eqnarray} 
where $n(P)$ is the number density of dust particle, $P$ is the ambient 
pressure, $P_0$ is the pressure at the base radius $R_P$ and $n_0$ is a
free parameter with dimension of number density. The deck of the
haze is fixed at 0.1 Pa  pressure level and the base is located at 1.5-2.5 Pa.
A log-normal size distribution of the dust particles given by
\begin{eqnarray}     
f(d)=\frac{d}{d_0} \times \exp\left[\frac{\ln(d/d_0)}{\ln \sigma}\right]^2
\end{eqnarray}
where $d$ is the diameter of the dust particle, $d_0$ is the median diameter
in the distribution and $\sigma$ is the standard deviation. Without loss of
generality, in the present model, we fix $\sigma=1.3$ and the fraction of 
the maximum amplitude of the distribution function at which we set the 
cutoff of the distribution is taken
to be 0.02. We have used the wavelength-dependent real and imaginary parts of
the refractive index for amorphous Forsterite ${\rm (Mg_2SiO_4)}$ which
is believed to be the dominant constituent of atmospheric cloud. 

   It must be emphasized that although cloud  or haze may play crucial role
in determining the transmission as well as the emission spectra of hot-Jupiters, it is not necessary that the atmosphere of all hot-Jupiters should
have cloud in the visible atmosphere. For low surface gravity and strong
irradiation, cloud may evaporate from the atmosphere. On the other hand, for
high surface gravity and low temperature, cloud may rain down
below the visible region.  The absence of alkaline absorption feature in 
the transmission spectra of many hot-Jupiters is usually interpreted as
the presence of cloud. 
The whole purpose of this work is to invoke additional absorption
and scattering opacities in the form of condensates and investigate how
the optical spectra are affected by dust (Mie) scattering over Rayleigh scattering.
In future, we shall incorporate more complicated and self-consistent cloud
models.

\section{Results and Discussions}

We present the results for wavelength region ranging from  near optical to
infrared.  Figure~5 shows the difference in the transmission depth 
calculated by solving the multiple scattering radiative
transfer equations and by using Beer-Bouguer-Lambert law.
The transmission depth presented by the model of \cite{kempton17} using 
Beer-Bouguer-Lambert law  overlaps at all wavelengths with that of our model
when $\omega=0$. Note that the opacity due to scattering is,
however, included in the calculations of the optical depth.         

\begin{figure}[]
\includegraphics[angle=-90,scale=0.7]{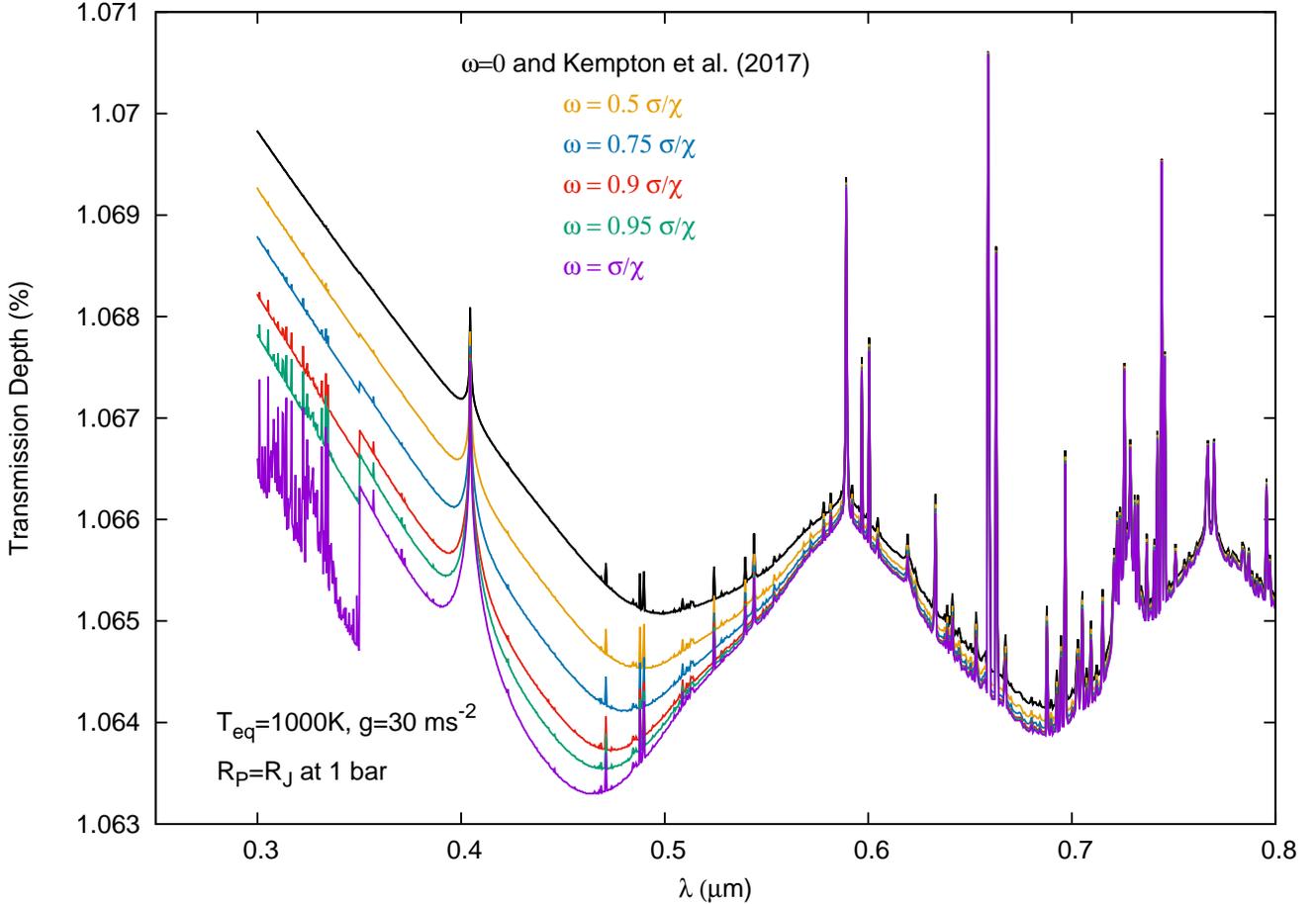}
\caption{Transmission depth of a hot-Jupiter for different values of
the scattering albedo $\omega=\sigma/\kappa$. The optical depth is unaltered for all cases.
The model by \cite{kempton17} adopts
Beer-Bouguer-Lambert law $I=I_0e^{-\tau}$ which overlaps with the present model
when $\omega$ is set at zero. 
All of our models use the solution of the radiative transfer equations.    
\label{fig:fig5}}
\end{figure}

\begin{figure}[]
\includegraphics[angle=-90,scale=0.7]{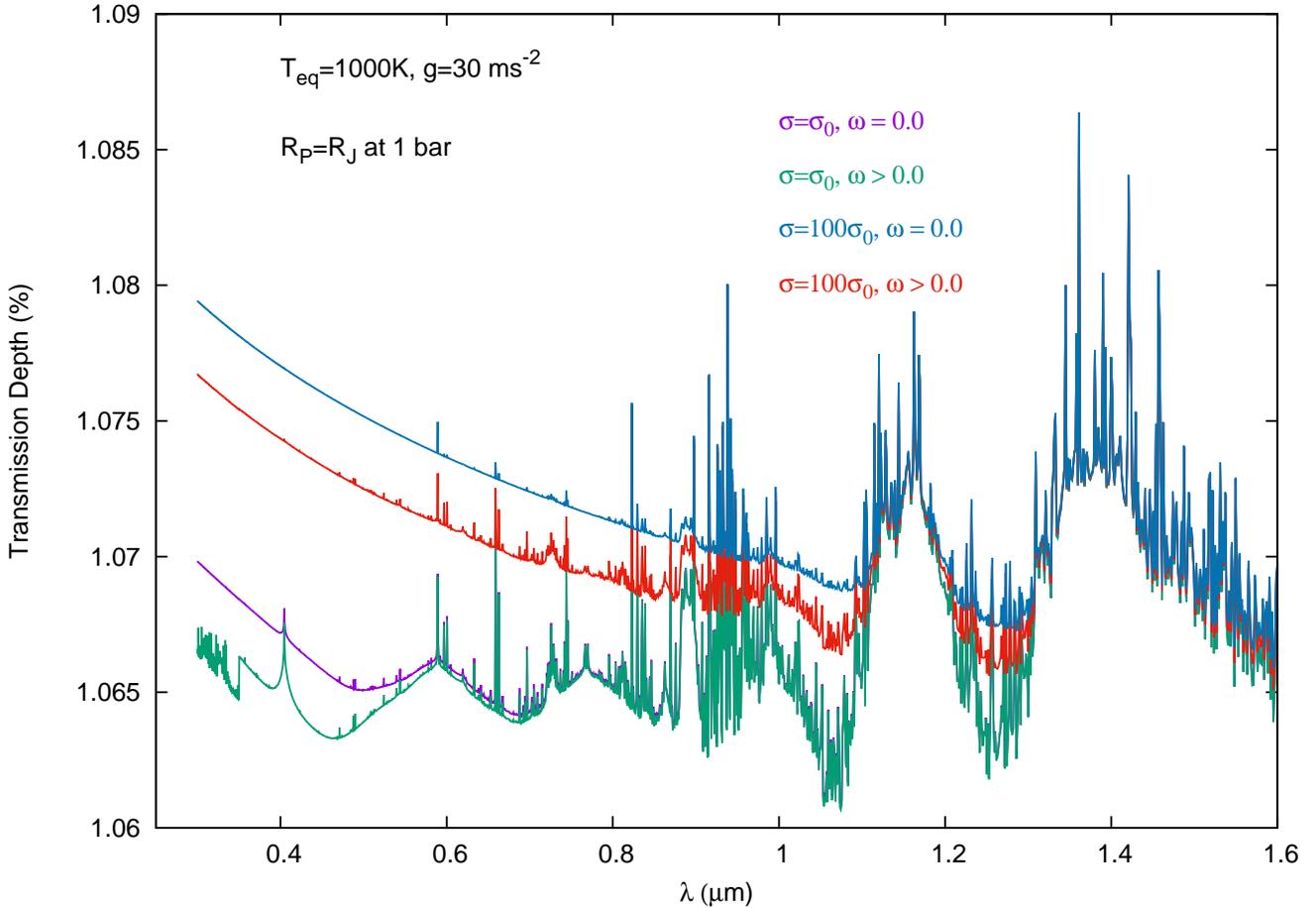}
\caption{Comparison of transmission spectra with increased scattering
opacity for zero and non-zero albedo.
$\sigma_0$ is the actual opacity due to Rayleigh scattering derived by using 
solar system abundances. $\omega$ is the corresponding scattering albedo. The
absorption co-efficients at all wavelengths are kept unaltered.
\label{fig:fig6}}
\end{figure}

Considering a Jupiter-type exoplanet with $T_{eq}=1000$K and surface gravity
$g=30 {\rm ms^{-2}}$, we investigate the effect of scattering albedo by 
increasing  its value while unaltering the optical depth due to scattering.
Figure~\ref{fig:fig5} shows that with the increase in the 
scattering albedo $\omega$, the amount of diffuse radiation due to scattering
increases. Part of this diffuse radiation is added to the reduced stellar light that
transmit the planetary atmosphere.
Consequently the transmitted flux increases amounting in a decrease in
the transmission depth.  However, at wavelength longer than
about 0.6 $\mu{\rm m}$, the Rayleigh scattering albedo becomes negligibly
small and therefore the transmission spectra coincide to that without scattering.
Hence, Figure~\ref{fig:fig5} demonstrates that scattering plays an important
role in determining the optical transmission spectra. Clearly, scattering 
contributes in two ways - (1) the opacity due to scattering adds up to the
opacity due to pure absorption and hence increases the total optical
depth which reduces the transmitted stellar flux and (2) increases the 
transmitted stellar flux by adding the diffuse radiation due to scattering 
to the outgoing stellar flux. The net effect yields into a decrease 
in the transmission depth as shown in Figure~5.  However, at about $0.35 {\rm \mu m}$,
we notice a sudden rise in the transmission depth which remains unexplained. A possible
reason could be a small but sharp increase in absorption by Na and K  in the opacity data used
which reduces the scattering albedo at that wavelength.
The effect of scattering becomes negligible at wavelengths longer than about 
$0.7{\rm \mu m}$ where the opacity due to scattering also becomes negligible. 
This is also demonstrated in Figure~\ref{fig:fig6}. However, with the increase in the scattering
co-efficients, the effect of scattering albedo is significant even in the near-infrared.   

Figure~7 demonstrates that the transmission depth increases if the stellar flux
passes through the deeper region of the atmosphere. If we consider that the
planetary atmosphere through which the stellar flux is transmitted is extended up to
a pressure level of 10 bar instead of 1 bar, the transmission depth increases by an
amount given in Equation~3. Note that, in that case both the first and the
second terms in the right hand side of Equation~3 should affect the transmission
depth. However, Figure~7 shows a constant difference in the transmission depth even
up to 10 $\mu{\rm m }$ implying that the change in the atmospheric radius $R_{PA}$ plays
dominant role over the change in the transmitted flux. 
However, as mentioned in Section~\ref{sec:tpprofile}, because of the transit geometry 
considered for calculating the transmission spectrum, the atmosphere below 
approximately 1 bar pressure level is sufficiently opaque to the transmitted stellar 
radiation. Therefore, in all models we calculate $R_{PA}$ at 1 bar pressure level of
the planetary atmosphere.

Similarly, Figure~8 shows that
the transmission spectra does not differ significantly
if isothermal temperature-pressure profile is considered instead of 
non-isothermal temperature-pressure profile derived through detail
numerical procedure.  However, as Figure~4 implies, the isothermal approximation
is reasonable only if the planet is not strongly irradiated. For planets with equilibrium
temperature higher than about 1400K, presence of TiO/VO introduces significant inversion
in the temperature and therefore even at the upper layer of the atmosphere
isothermal approximation may not be appropriate.
Therefore, in order to calculate the transmission 
spectra in the optical region, accurate non-isothermal temperature-pressure
profiles are needed to be used for relatively hotter planets.

\begin{figure}[]
\includegraphics[angle=-90,scale=0.7]{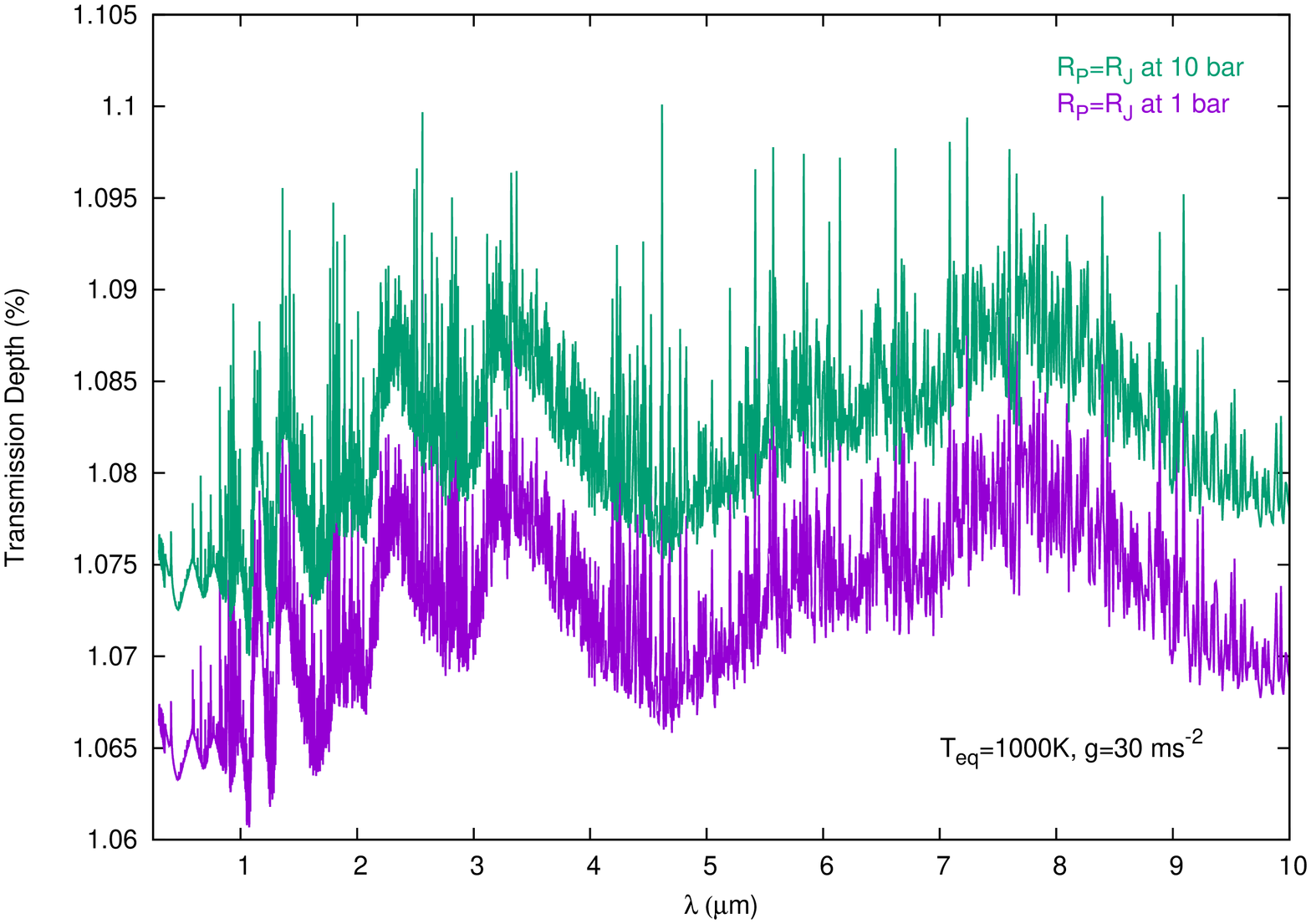}
\caption{Comparison of transmission spectra with the base radius $R_P$ located
at different pressure levels.
\label{fig:fig7}}
\end{figure}

\begin{figure}[]
\includegraphics[angle=-90,scale=0.7]{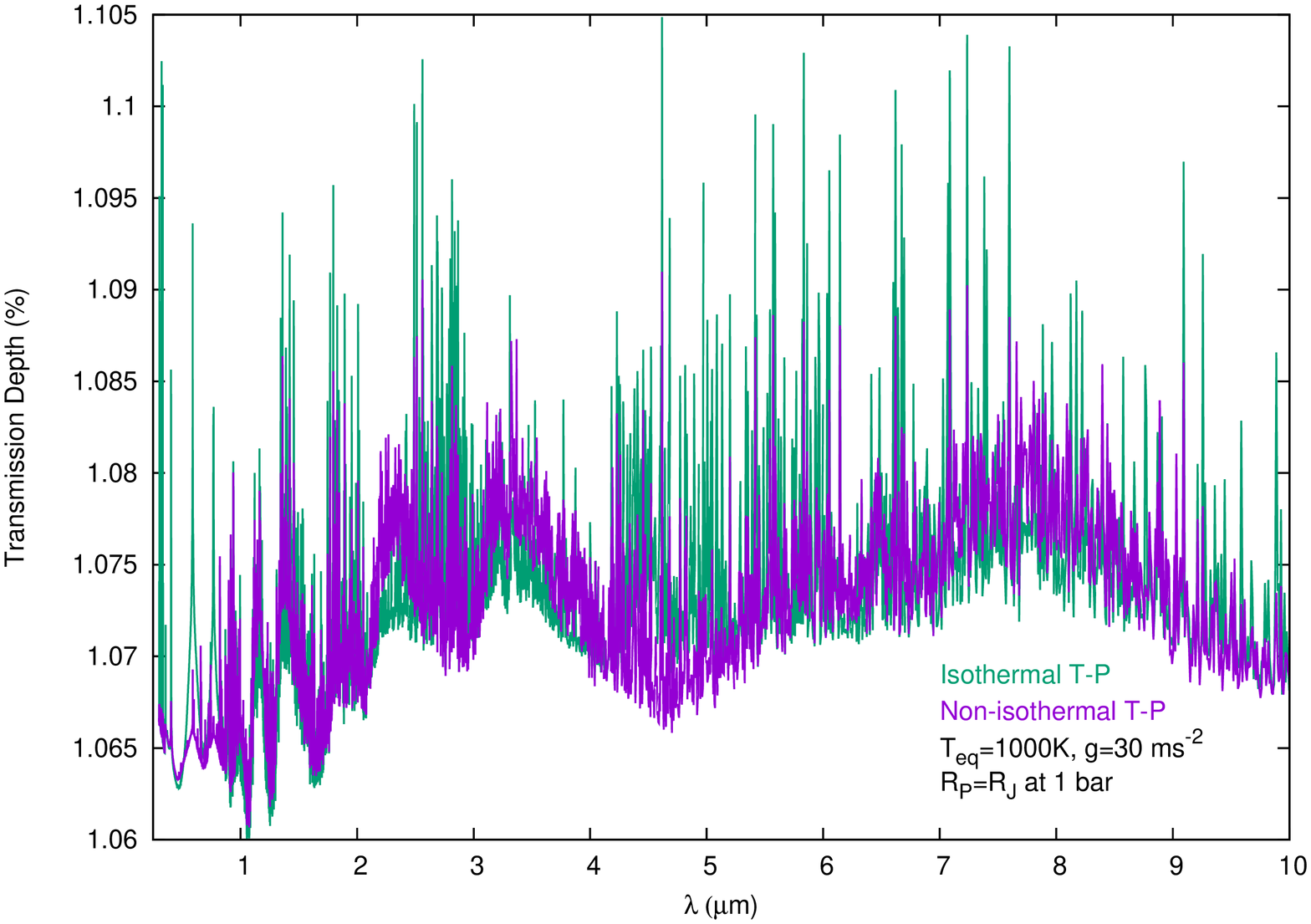}
\caption{Comparison of transmission spectra with isothermal and non-isothermal
temperature-pressure profile. For isorthermal case, the temperature is taken to
be equal to $T_{eq}$ at all pressure points. 
\label{fig:fig8}}
\end{figure}

 Using non-isothermal temperature-pressure profiles, we calculated the 
absorption and scattering coefficients and then the transmission 
depth is calculated by solving multiple scattering radiative transfer equations for 
plane-parallel stratification of the planetary atmosphere. We compare our
model spectra for a few hot-Jupiters with the existing model spectra and
observed data presented by \cite{sing16}. The model spectra of
\cite{sing16} and the observed data are available in public domain\footnote{ 
${\rm https://pages.jh.edu/~dsing3/David_Sing/Spectral_Library.html}$}. 
The detail about these model grids and
about the observed data are described in \cite{sing16}. The various physical
parameters adopted in order to obtain best fit (by eye) at the infra-red region where
scattering is negligible are listed in Table~1.

\begin{figure}[]
\includegraphics[angle=-90,scale=0.7]{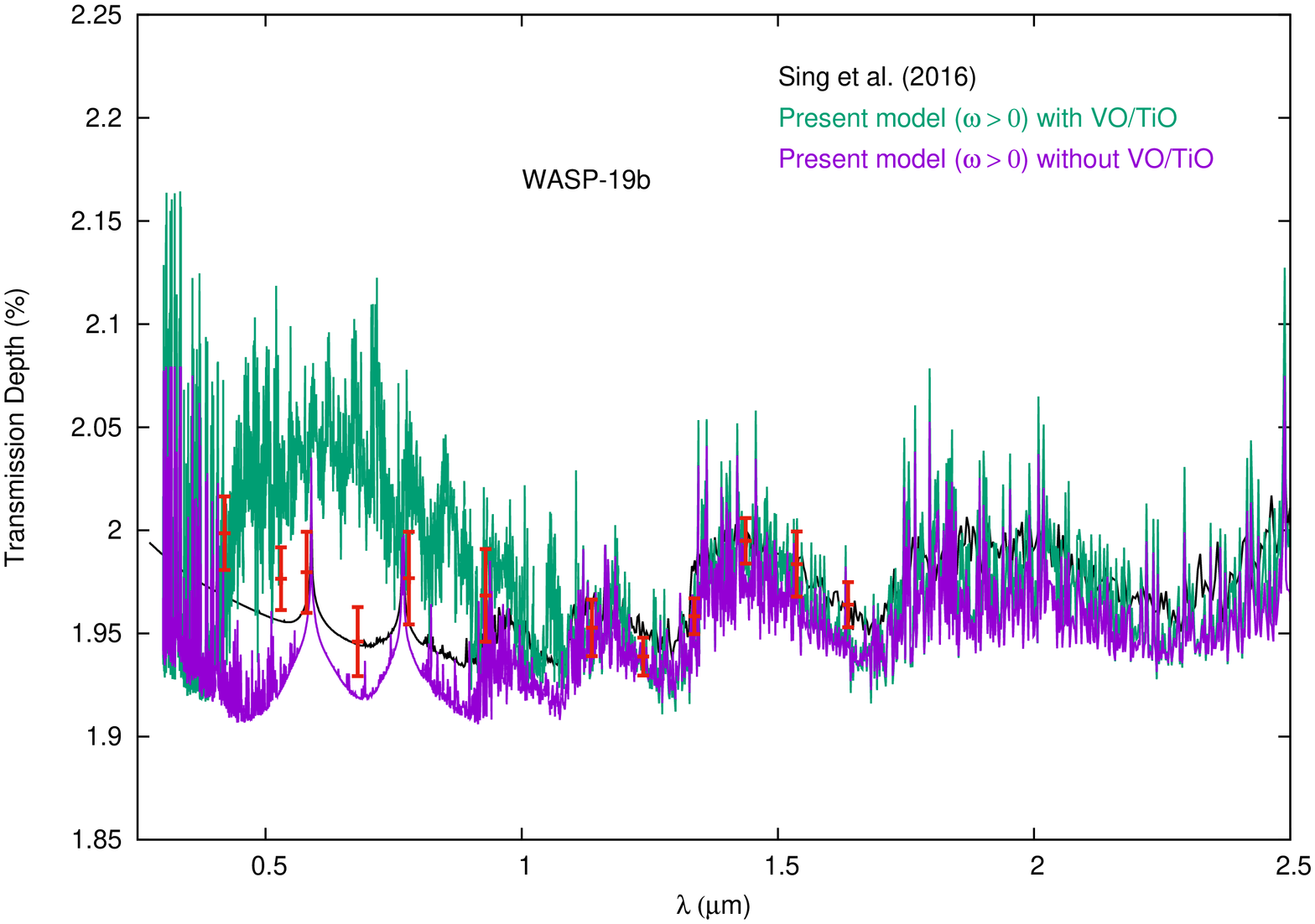}
\caption{ Comparison of model transmission spectra with and without the presence of VO and
TiO and with the observed data (red) for WASP-19b.
\label{fig:fig9}}
\end{figure}

  In order to explain the high transmission depth in the optical, \cite{sing16}
incorporated additional Rayleigh scattering opacity by increasing the scattering
cross section of hydrogen molecules 10 to 1000 times its value at 350 nm.. 
 However, as can be seen in Figure~9, the present model for the
exoplanet WASP-19b yields much higher transmission depth at optical region 
up to 1.0 $\mu{\rm m}$ than that presented by \cite{sing16}. We have not included any additional
opacity source for this model. This difference is attributed to the presence of TiO and VO.
In the absence of TiO and VO, the transmission depth profile qualitatively matches well
with the model by \cite{sing16}. However, since diffusion by scattering reduces the transmission
depth, the two models do not overlaps. The two models, however, match at wavelengths
longer than 1.0 $\mu{\rm m}$ where the scattering is negligible.. 

  As mentioned before, formation of cloud in the planetary atmosphere needs
appropriate combination of temperature and surface gravity. Strong irradiation
or strong thermal radiation can cause evaporation of the cloud while low
temperature and high  surface gravity may cause rain out of the condensates.
The disappearance of atomic and molecular absorption lines in the optical is
usually interpreted as the evidence of cloud or haze. However, for planetary
atmosphere that has no or negligible thermal radiation, scattering by cloud may alter 
the absorption features in the transmission spectra. Presence of cloud or
haze not only changes the total opacity of the atmosphere, it also alters the
scattering albedo of the medium..

 For all models except that of WASP-19b, we have included a thin haze
in the upper atmosphere as described in section~\ref{sec: cloud}. In Figure~10,
we present a comparison of the transmission spectra for HD 209458b
with and without haze. We also present in the same figure the transmission
spectra obtained with $\omega=0$. For both the cases - with zero and 
non-zero albedo, the total extinction i.e., the opacity due to true absorption
as well as scattering is kept unchanged. With the inclusion of haze, the
extinction increases yielding into higher optical depth at the upper 
atmosphere. The scattering albedo also changes due to cloud particles.
Figure~10 shows that the transmission depth calculated with or without Rayleigh
scattering albedo is much lower than that presented by \cite{sing16}. But a reasonably
good match with the model by \cite{sing16} in the optical is obtained by the 
inclusion of haze. The model spectra with and without haze converges at wavelength 
longer than 1.3 $\mu{\rm m}$. For this case, we have not presented the observed
data as it fits with the model by \cite{sing16} which fits the observed data in the optical.

  Similarly, we have obtained reasonably
good match with the model by \cite{sing16} as well as with the observed data for
HAT-P-1b by invoking haze in the planetary atmosphere. A comparison is presented
in Figure~11. The transmission depth calculated without haze is significantly less than that
calculated with haze. Note that in both the cases, the effect of scattering albedo is
included. All the model spectra, however converges  
at wavelength longer than about 1.3 $\mu{\rm m}$ where scattering becomes negligible..

\begin{figure}[]
\includegraphics[angle=-90,scale=0.7]{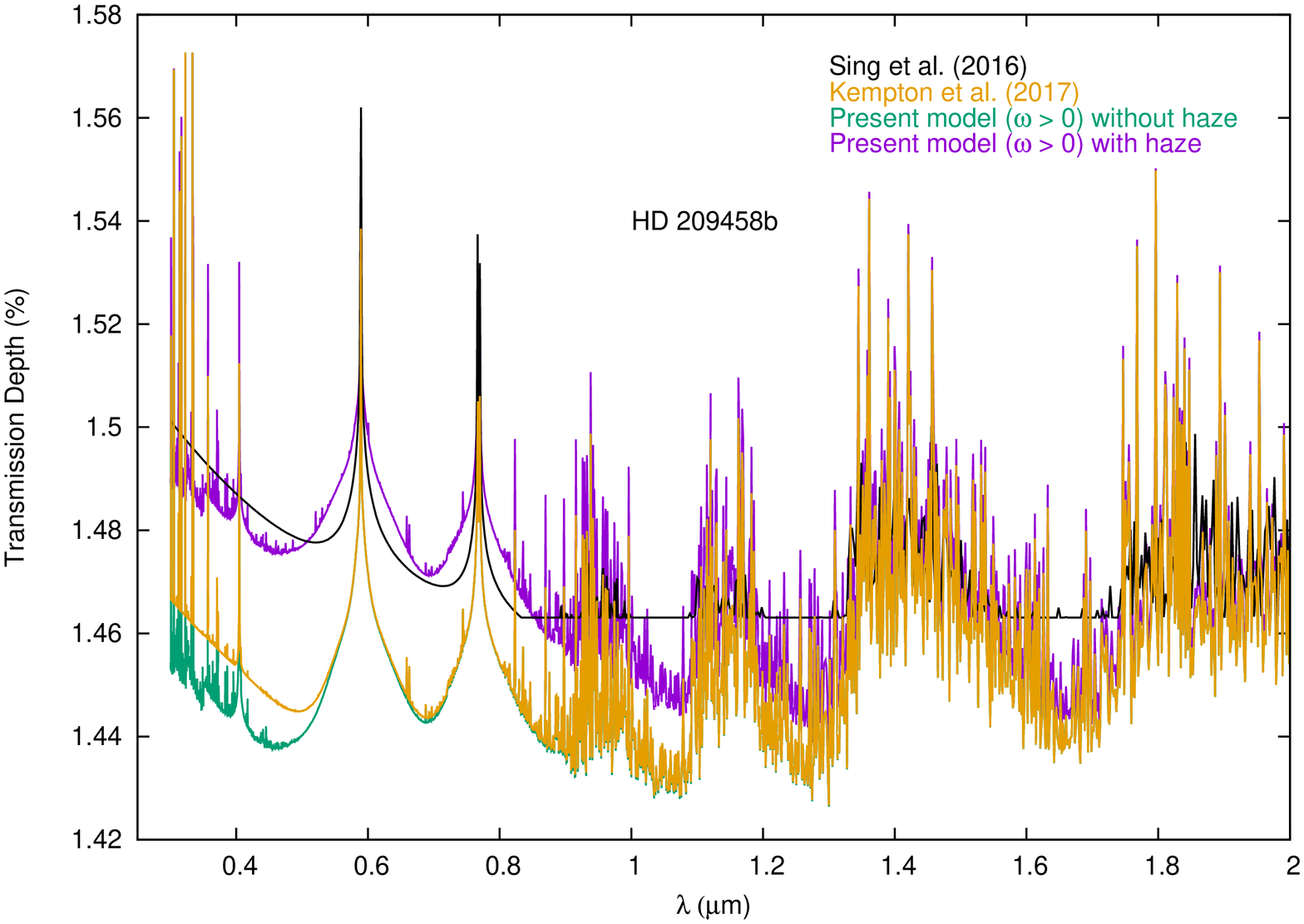}
\caption{Comparison of model transmission spectra with and without the effect of
Rayleigh scattering albedo and that by haze for exoplanet HD 209458b.
\label{fig:fig10}}
\end{figure}

\begin{figure}[]
\includegraphics[angle=-90,scale=0.7]{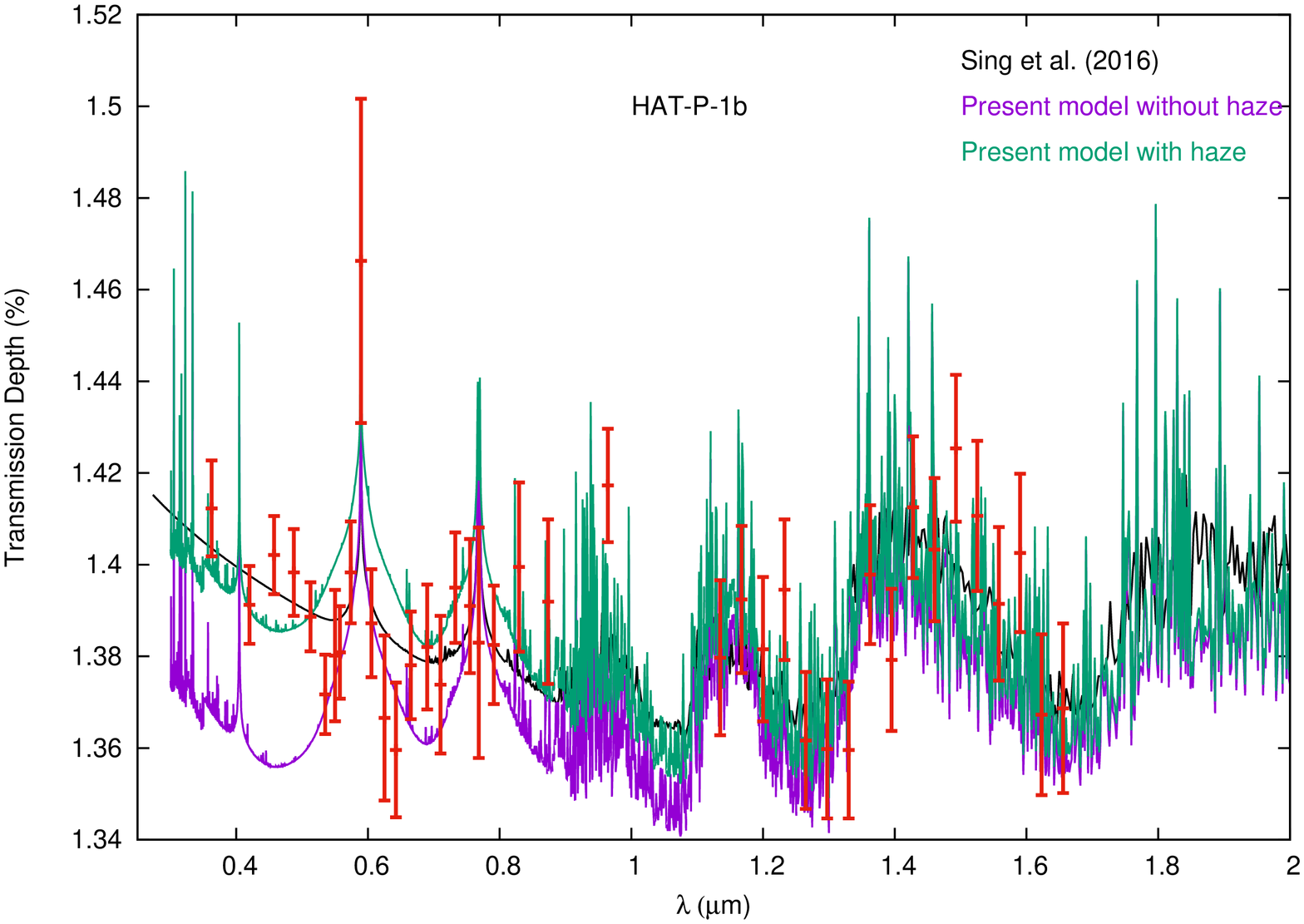}
\caption{Comparison of model transmission spectra with and without the effect of
haze and with the observed data (red) for HAT-P-1b. 
\label{fig:fig11}}
\end{figure}

  The sharp increase in the values of observed transmission depth for HD 189733b
and HAT-P-12b at wavelength shorter than 1.0 $\mu{\rm m}$ however do not fit our
model transmission spectra even by increasing the scattering opacity a thousand 
times or by incorporating haze. Figure~12, however, demonstrates that
inclusion of sub-micron size haze can produce comparable transmission spectrum  
that is obtained by invoking additional Rayleigh scattering opacity in the 
model by \cite{kempton17}. Figure~13 also demonstrate the difference in the transmission
spectra with and without the effect of scattering albedo when the scattering opacity is
increased by a thousand times. Inclusion of haze results into transmission depth comparable
to that presented by \cite{sing16} in the optical only if the diffusion by scattering 
is excluded in the model. Clearly, the diffuse radiation due to scattering increases
the transmitted flux resulting into a decrease in the transmission depth even up to 2.0
$\mu{\rm m}$. However, all the models converge at wavelengths longer than 
2.0 $\mu{\rm m}$ as the scattering coefficient and hence the scattering albedo becomes
negligible beyond this wavelength. We point out here that the rapid increase in the
transmission depth at wavelength shorter than 0.5 micron may be due to other
effects, e.g., stellar activities, star-spots etc.    

\begin{figure}[]
\includegraphics[angle=-90,scale=0.7]{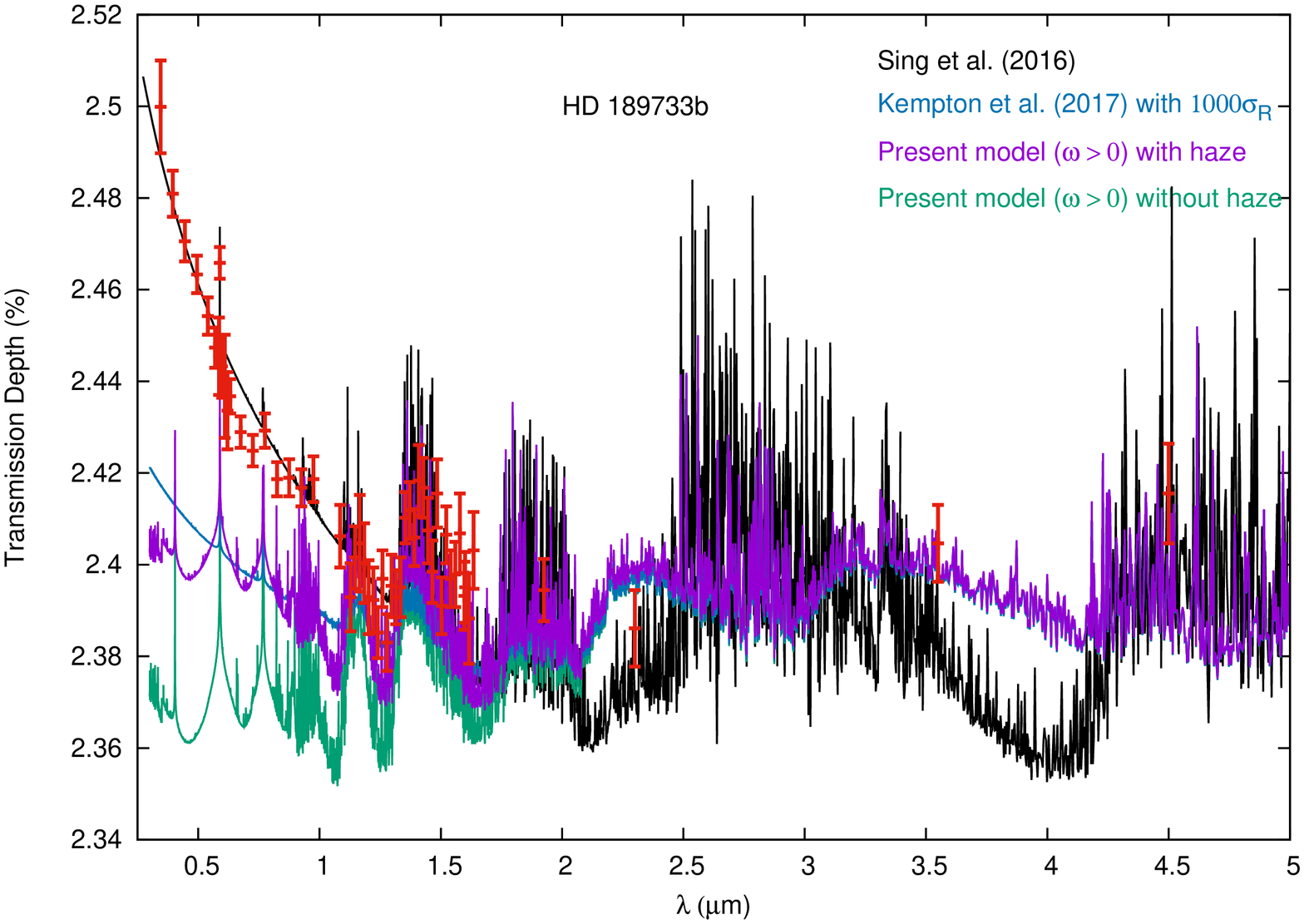}
\caption{Comparison of observed data (red) and model transmission
spectra for HD 189733b with and without haze in the upper atmosphere.
\label{fig:fig12}}
\end{figure}

\begin{figure}[]
\includegraphics[angle=-90,scale=0.7]{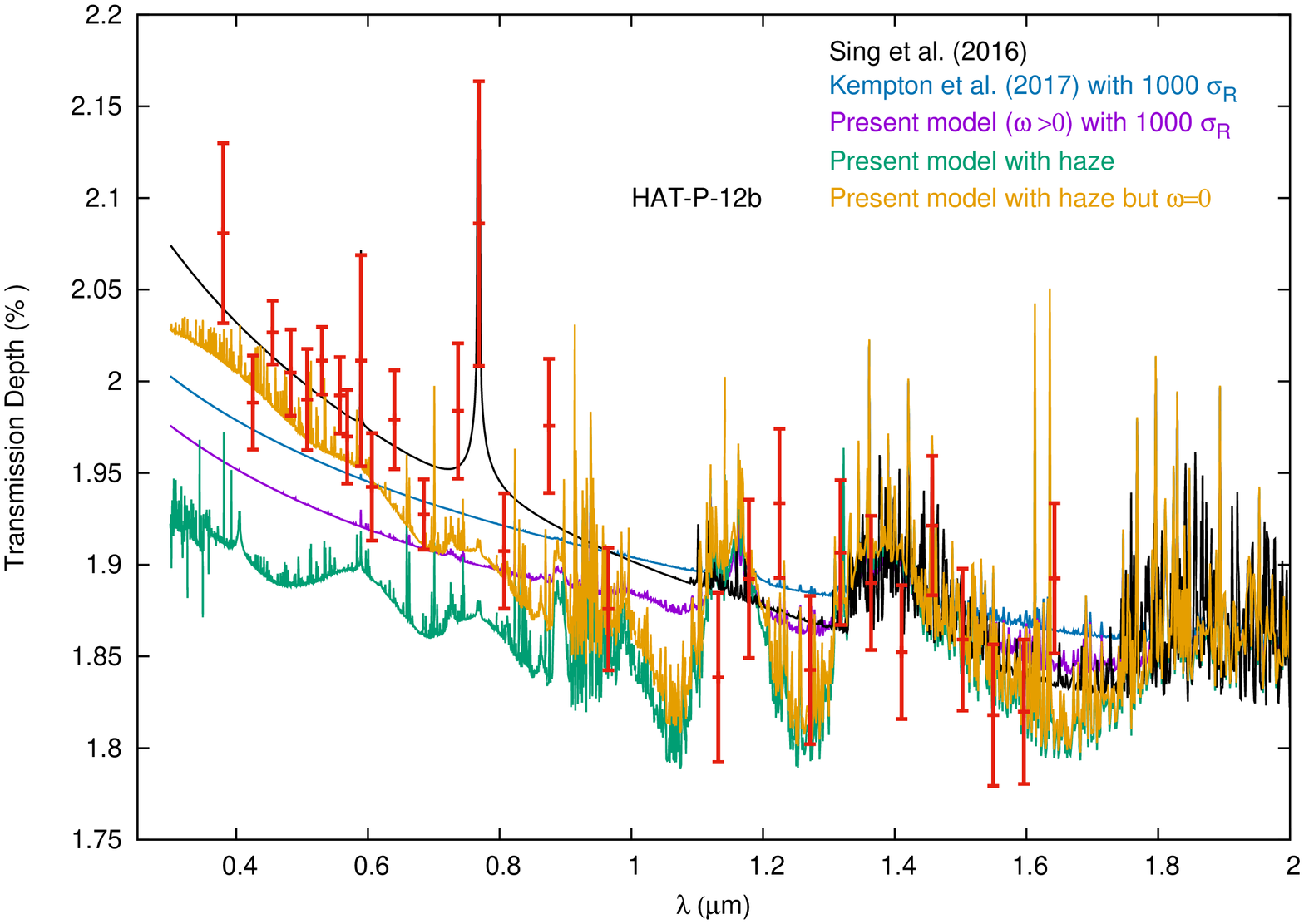}
\caption{Comparison of observed data (red) and model transmission
spectra for HATP-P-12b  with and without haze in the upper atmosphere. The model
transmission spectrum of \cite{kempton17} with thousand times of the actual scattering
co-efficient is also presented for a comparison.  Further, an atmospheric model with 
absorption by haze but without the effect of scattering albedo is presented in this figure. 
\label{fig:fig13}}
\end{figure}

 Finally, we present in Figure~14 the model transmission spectra for WASP-6b with 
and without incorporating haze. It is worth mentioning that our numerical method ensures 
that the dust number density does not exceed the mass of heavy elements. Figure~14
shows that even with the maximum allowed values of dust number density, the transmission
depth fails to fit the observed data in the optical region. We have achieved a good model fit
with the observed data by increasing the Rayleigh scattering opacity eight times its original
value in addition to incorporating haze. This indicates that a better cloud model is needed
to fit the observed data. 

\begin{figure}[]
\includegraphics[angle=-90,scale=0.7]{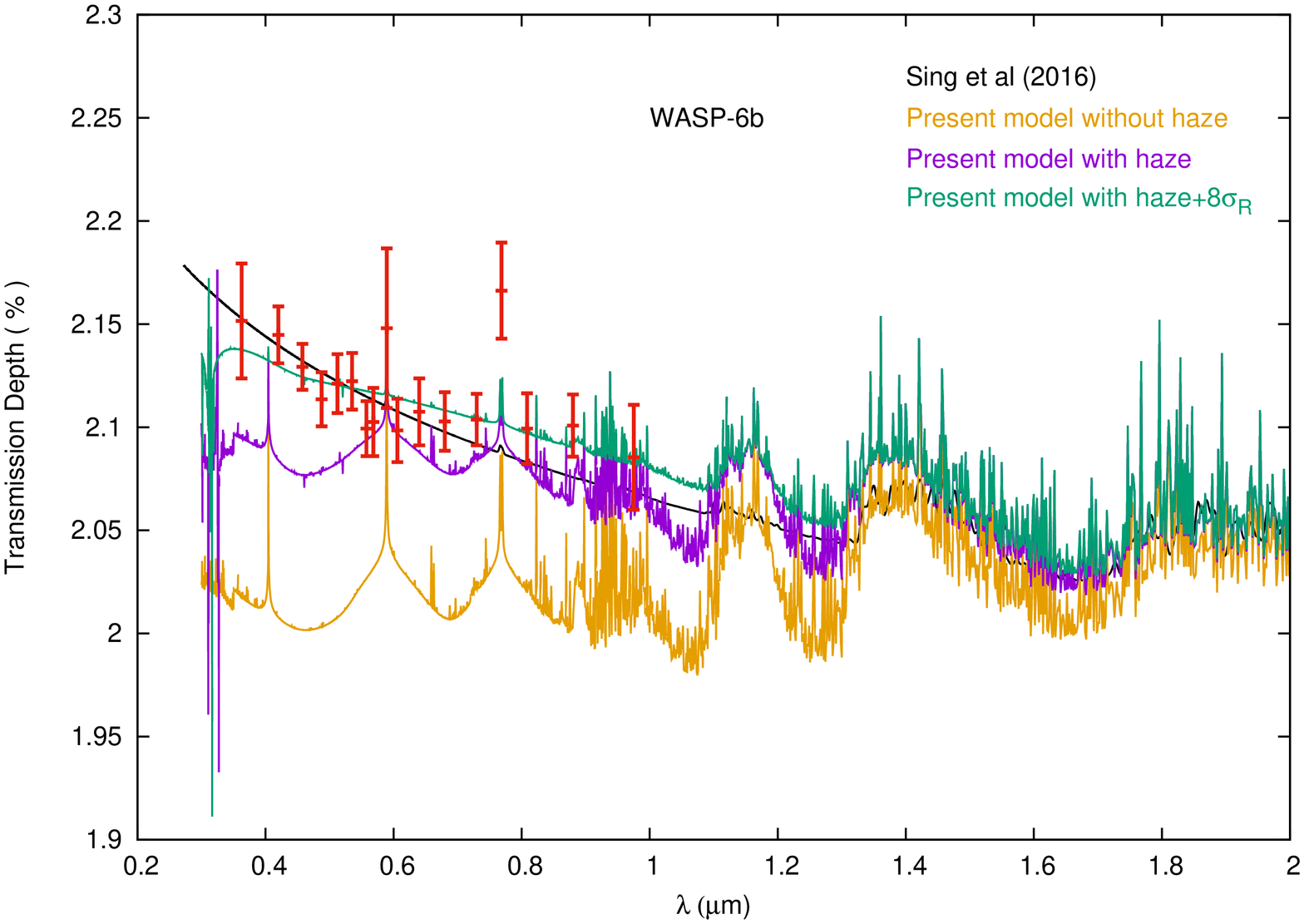}
\caption{Comparison of observed data (red) and model transmission
spectra for WASP-6b with and without haze. 
\label{fig:fig14}}
\end{figure}

\section{Conclusions}

  We have presented detail numerical models of transmission spectra for
hot jupiter-like exoplanets by solving the multiple scattering radiative
transfer equations with non-zero scattering albedo instead of using the
Beer-Bouguer-Lambert law. We have demonstrated that the solution
of the radiative transfer equations that incorporate the diffuse reflection
and transmission radiation field due to scattering yields significant 
changes in the transmission depth at the optical wavelength region, specially if
the atmosphere is cloudy. 
However, at longer wavelength scattering becomes negligible and the
transmission spectra overlap
with that derived by using Beer-Bouguer-Lambert law. We compare our model
spectra with the observed data and with two 
different theoretical models that include opacity due to scattering but
do not take into account the diffuse reflection and transmission of the
incident radiation field due to atmospheric scattering. We also include additional
opacity and scattering albedo due to condensate cloud by adopting a simplified
dust model. The most important message conveyed by the present work is that
in order to analyze the observed optical transmission spectra of exoplanets,
the retrieval models need to incorporate the scattering albedo that gives rise to
diffused radiation field which is added to the stellar radiation transiting through
the planetary atmosphere. Thus a correct and consistent procedure is to solve
the multiple scattering radiative transfer
equations. A substantial amount of diffuse stellar radiation
increases the transmitted flux resulting into a decrease in the transmission
depth. However, in the infrared wavelength region where the affect of scattering is 
negligible, Beer-Bouguer-Lambert law can very well be employed to calculate the transmission
depth.     

\acknowledgments

This project has received funding from the European Union's Horizon 2020 
research and innovation programme 776403, ExoplANETS A  and the Science
 and Technology Funding Council (STFC) grants: ST/K502406/1 and ST/P000282/1.
SS and AC would like to thank University College London, UK for hospitality
during their visit. We thank the reviewer for a critical reading and several
valuable comments and suggestions that enabled significant improvement of the
work. 

\software{Exo\_Transmit \citep{kempton17}, TauREx \citep{waldmann15}, Analytical$\;$model$\;$for$\;$irradiated$\;$atmosphere
 \citep{parmentier14,parmentier15}}


\begin{thebibliography}{}

\bibitem[Ackerman \& Marley (2001)]{ackerman01}  Ackerman, A. \&  Marley, M. S.
2001, ApJ, 556, 872.
\bibitem[Asplund, Grevesse, Sauval \& Scott (2009)]{asplund09}
Asplund M., Grevesse N., Sauval A. J., and Scott P., 2009, ARA\&A, 47, 481. 
\bibitem[Barstow, Aigrain, Irwin et al., (2017)]{barstow17} J. K. Barstow et al., 2017, ApJ, 834, 50.
\bibitem[Bohren \& Huffman (1983)]{bohren83} Bohren, C. F., \& Hu†man, D. R.
1983, Absorption and Scattering of Light by Small Particles (Wiley : New York).
\bibitem[Brown (2001)]{brown01} Brown, T.,  ApJ, 553:1006-1026, 2001
\bibitem[Burgasser, McElwain, Kirkpatrick et al. (2004)]{burg04} 
Burgasser, A. J., McElwain, M. W.., Kirkpatrick, J. D. et al. 2004, AJ , 127,
2856.  
\bibitem[Burrows, Budaj \& Hubeny (2008)]{burrows08} Burrows, A., Budaj, J.
\& Hubeny, I. 2008, \apj, 678, 1436.
\bibitem[Burrows et al. (2010)]{burrows10}  A. Burrows et al. 2010, ApJ, 719, 341.
\bibitem[Burrows, 2014]{burrows14}  Burrows, A. S., 2014, PNAS, 111 (35) 12601-12609
\bibitem[Chandrasekhar (1960)]{chandra60} Chandrasekhar, S. Radiative
Transfer (New York: Dover, 1960).
\bibitem[Cooper, Sudarsky,Milsom, et al.(2003)]{cooper03} Cooper, C. S., 
Sudarsky, D., Milsom, J. A., Lunine, J. I. \& Burrows, A.  2003, \apj,
586, 1320.
\bibitem[Cushing, Marley, Saumon et al. (2008)]{cushing08} Cushing, M. C., Marley, M. S., Saumon, D.  et al. 2008, 
ApJ, 678, 1372.
\bibitem[de Kok \& Stam (2012)]{stam2012} de Kok, R. J. \& Stam, D. M. 2012, Icarus, 221, 517.
\bibitem[Fortney, Marley, Saumon et al. (2008)]{fortney08} Fortney, J. J.,
Marley, M. S., Saumon, D. \& Lodder, K. 2008, \apj, 683, 1104. 
\bibitem [Fortney, Shabram, Showman et al. (2010)]{fortney10} Fortney, J. J.,
Shabram, M., Showman, A. P., et al. 2010, \apj, 709, 1396.
\bibitem[Fortney (2018)]{fortney18} Fortney, J. J. 2018, eprint
arXiv:1804.08149. 
\bibitem [Freedman, Marley \& Lodder (2008)]{freedman08} Freedman, R. S., 
Marley, M. S., \& Lodders, K. 2008, ApJS, 174, 504. 
\bibitem[Freedman, Lustig-Yaeger, Fortney et al. (2014)]{freedman14} Freedman, 
R. S., Lustig-Yaeger, J., Fortney, J. J., et al., 2014, ApJS, 214, 25.
\bibitem[Goyal, Mayne, Sing et al. (2018)]{goyal18} Goyal J. M., Mayne, N. J.,
Sing, D. K.,  et al., 2018, MNRAS, 474, 5158.
\bibitem[Goyal, Wakeford, Mayne et al. (2019)]{goyal19} Goyal, J. M., Wakeford,
H. R., Mayne, N. J. et al. 2019, MNRAS, 482, 4503. 
\bibitem[Gordon et al. (2017)]{gordon17} I.E. Gordon, L.S. Rothman, C. Hill et al., 2017, J Quant Spectrosc Radiat Transfer 203, 3-69.
\bibitem[Guillot (2010)]{guillot10} Guillot, T. 2010, A\&A, 520, A27.
\bibitem[Griffith, Yelle, \& Marley (1998)]{griffith98} Griffith, C. A., Yelle, R. A., \&  Marley, M. S. 1998, Science, 282, 2063. 
\bibitem[Griffith (2014)]{griffith14} Griffith, C. A., 201, 372, Philosophical Transactions of the Royal Society A: Mathematical, Physical and Engineering Sciences.
\bibitem[Hansen (2008)]{hansen08} Hansen, B. M. S. 2008, \apjs, 179, 484. 
\bibitem[Heng  et al. (2018)]{heng18} K. Heng et al 2018, ApJS, 237, 29.
\bibitem [Kempton, Lupu, Owusu-Asare et al.  (2017)]{kempton17} Kempton, 
E. M.-R., Lupu, R. E., Owusu-Asare, A., Slough, P., \& Cale, B., 2016, PASP,
129, 1 
\bibitem[Lodders, K. (2003)]{lodders03} Lodders, K. 2003, \apj, 591, 1220. 
\bibitem[Lupu, Zahnle, Marley et al. (2014)]{lupu14} Lupu, R. E., Zahnle, K., 
Marley, M. S., et al., 2014, ApJ, 784, 27 
\bibitem[Madhusudhan and Seager (2009)]{madhu09} N. Madhusudhan and S. Seager, 2009, ApJ, 707, 24
\bibitem[Marley \& Sengupta (2011)]{marley11} Marley, M. S. \& Sengupta, S. 
2011, MNRAS,417, 2874.
\bibitem [Parmentier \& Guillot (2014)]{parmentier14} Parmentier, V. 
\& Guillot, T. 2014, A\&A, 562, A133.
\bibitem[Parmentier, Guillot, Fortney et al. (2015)]{parmentier15} Parmentier,
V., Guillot, T., Fortney, J. J., \& Marley, M. S. 2015, A\&A, 574, A35. 
\bibitem[Peraiah \& Grant (1973)]{peraiah73} Peraiah, A., \& Grant,
I. P.  1973, J. Inst. Maths. Appl. 12, 75.
\bibitem[Saumon, Geballe, Leggett, et al. (2000)]{saumon00} Saumon, D.,
Geballe, T. R., Leggett, S. K., et al. (2000), \apj, 541, 374. 
\bibitem[Seager \& Sasselov (2000)]{seager00} Seager S., \& Sasselov D. D.,
2000, ApJ, 537, 916.
\bibitem[Sengupta \& Marley (2009)]{sengupta09} Sengupta, S. \& Marley, 
M. S. 2009, ApJ, 707, 716.
 \bibitem[Sengupta \& Marley (2010)]{sengupta10} Sengupta, S. \& Marley, 
M. S. 2010, ApJL, 722, L142.
\bibitem[Sengupta \& Marley (2016)]{sengupta16} Sengupta, S. \& Marley, 
M. S. 2016, ApJ, 824, 76..
\bibitem[Sengupta (2016)]{sengupta16a} Sengupta, S. 2016, AJ, 152, 98.
\bibitem[Sengupta (2018)]{sengupta18} Sengupta, S. 2018, \apj, 861, 41. 
\bibitem[Sing, Fortney, Nikolov et al. (2016)]{sing16} Sing, D. K., Fortney,
J. J., Nikolov, N. et al. 2016, Nature, 529, 59.   
\bibitem[Sudarsky, Burrows \& Hubeny (2003)]{sudars03} Sudarsky, D., Burrows,
 A., \& Hubeny, I. 2003, \apj, 588, 1121.
\bibitem[Stephens et al. (2009)]{stephens09} Stephens, D. C. et al. 2009, ApJ, 702,154.
\bibitem[Sudarsky, Burrows \& Hubeny (2003)]{sudarsky03} Sudarsky, D., Burrows,
A., \& Hubeny, I. 2003, \apj, 588, 1121.
\bibitem[Tennyson \& Yurchenko (2012)]{tennyson12} Tennyson, J. \& Yuschenko,
S. N., 2012, MNRAS, 425, 21. 
\bibitem[Tennyson, Yurchenko, Al-Refaie, et al. (2016)]{tennyson16} 
Tennyson, J., Yurchenko, S. N., Al-Refaie, A. F., et al. 2016, J. Molecular
 Spectroscopy, 327, 73.
\bibitem[Tinetti, Encrenaz \& Coustenis (2013)]{tinetti13} Tinetti, G., Encrenaz, T. \&  Coustenis, A. 2013, AAR, 21, 63.
\bibitem[Tinetti,  Liang, et al. (2007)]{tinetti07}  Tinetti, G., M. C. Liang et al., 2007, ApJ, 654, L99.
\bibitem[Tsiaras, Waldmann, et al. (2018)]{tsiaras18} A. Tsiaras, I. P. Waldmann et al 2018, AJ, 155 156.
\bibitem[Valencia, Guillot, Parmentier et al (2013)]{valencia13} Valencia, D.,
Guillot, T.,Parmentier, V. \& Freedman, R. S. 2013, \apj, 775, 10. 
\bibitem[Waldmann, Tinetti, Rocchetto, et al. (2015)]{waldmann15} 
Waldmann, I. P., Tinetti, G., Rocchetto, M. et al. 2015, \apj, 802, 107.
\bibitem[Yip et al. (2019)]{yip19} Yip K. H. et al., 2019, Integrating light-curve and atmospheric modelling of transiting exoplanets, arXiv:1811.04686.


\end{thebibliography}
\end{document}